\documentclass[prb,twocolumn,showpacs,amsmath,amssymb,aps]{revtex4}
\usepackage{graphicx}% Include figure files
\usepackage{dcolumn}% Align table columns on decimal point
\usepackage{bm}% bold math

\begin{document}

%\preprint{APS/123-QED}

\title{Inverse Tunneling Magnetoresistance in nanoscale Magnetic Tunnel Junctions}
\author{Tae-Suk Kim}
\affiliation{School of Physics, Seoul National University,
  Seoul 151-747, Korea} 
\date{\today}

\begin{abstract}
 We report on our theoretical study of the inverse TMR effect in the spin polarized transport 
through a narrow channel. In the weak tunneling limit, we find the ordinary 
positive TMR. The TMR changes its sign as the transmission probability becomes large
close to a unity. Our results might be relevant to the magnetic tunnel junction with a pinhole
or a quantum point contact.

\end{abstract}
\pacs{72.25.-b, 73.40.Gk}
\maketitle

\section{Introduction}
 Recently spin-polarized transport has attracted lots of attention because of its
potential applications to spin-electronic devices. 
Magnetic tunnel junction (MTJ),\cite{review,review2} consisting of two ferromagnetic (FM) electrodes 
with an insulating layer sandwiched in between two, is one exemplary realization 
of the spin-polarized transport and show a large tunneling magnetoresistance (TMR)
effect.\cite{moodera,miyazaki} Stable MTJs with large TMR are now routinely fabricated.
%their possible applications as magnetic memory cells and sensors. 
The tunneling current is modulated by the relative orientation of 
magnetizations in the two FM electrodes. 
According to the Julliere's model \cite{julliere}, the TMR ratio is given by 
\begin{eqnarray}
TMR &=& \frac{R_{AP} - R_P} {R_P} ~=~ \frac{2 P_L P_R}{1 - P_L P_R}. 
\end{eqnarray}
Here $R_P$ and $R_{AP}$ are the tunnel resistance for the parallel and antiparallel 
alignment of the MTJ, respectively, and $P_L$ and $P_R$ are the electron spin 
polarizations of two FM electrodes.  
Usually more current flows in the parallel alignment of magnetization 
such that the TMR is positive. 
%The TMR is known to depend sensitively on the electronic states near the interface 
%between the ferromagnetic metal and the oxide barrier. 

 Recently some experimental groups \cite{teresa1, teresa2, sharma, itmr_tsymbal, fdotf_exp}
observed the inverse TMR effect for which 
more current flows in the antiparallel alignment of magnetizations in two FM electrodes 
than in the parallel alignment. 
The inverse TMR in Co/SrTiO$_3$/La$_{0.7}$Sr$_{0.3}$MnO$_3$ is ascribed \cite{teresa1} to
the selective tunneling of $d$ electrons of Co through the insulating SrTiO$_3$ barrier. 
To investigate the role of the oxide barrier on the sign of TMR, MTJs with various oxide 
barriers were studied. \cite{teresa2}  
Theoretical studies \cite{sdos_wang, sdos_tsymbal, sdos_butler} elucidated the important role 
of the surface DOS and the electronic wave functions in the barrier in determining 
the nature of tunneling electrons or the sign of the TMR.  
The surface density of states for Co and Fe near the Fermi energy is dominated 
by the $d$-band, and the $s(d)$-band in Co has the positive (negative) 
spin polarization (SP), respectively.\cite{sdos_wang, sdos_tsymbal, sdos_butler}
When the Al oxide layer is sandwiched between two Co or Fe metals,
the $s$-band electrons relatively easily tunnel through the barrier
while the $d$ electrons are strongly suppressed in the barrier.
The $s$ electrons are mainly responsible for the tunneling leading to the positive SP
\cite{sdos_wang, sdos_tsymbal, sdos_butler}.
For the barrier of SrTiO$_3$, the opposite is true, leading to the negative TMR 
or inverse TMR.

 The inverse TMR was also observed \cite{itmr_tsymbal} when the spin-polarized 
electrons tunnel resonantly through the impurity states in the barrier. 
More current in the antiparallel alignment can flow resonantly through a localized 
state in the barrier when the impurity is located asymmetrically inside the barrier
with respect to two FM electrodes. 
In this case the inverse TMR is determined mainly by the structure of the impurity states 
in the barrier. %irrespective of the bulk FM electrodes. 

 The inverse TMR can occur due to the Kondo effect when the impurity spin is sandwiched 
in between two FM electrodes.\cite{fdotf_the} 
The Kondo resonance peak can develop even when the magnetic 
impurity is coupled to the spin-polarized electrons of ferromagnetic metals. 
In the antiparallel alignment of two FMs, the impurity states of spin up and spin down 
are more symmetrically coupled to the corresponding spin states of conduction electrons 
than in the parallel alignment. 
When the coupling constants of both spin states are more or less equal, 
the whole Kondo resonance peak develops at the impurity site.\cite{fdotf_exp}
Either in the case of asymmetrical couplings between two FMs for the antiparallel 
configuration or in the parallel configuration, the Kondo spin states are split 
by the exchange field of two FMs such that the Kondo resonance peak is 
split into two. \cite{fdotf_the,fdotf_exp} 
Since the Kondo resonant state acts as the electric current channel, more current 
can flow in the antiparallel alignment than in the parallel alignment, resulting in 
the negative value of the TMR. The inverse TMR was indeed observed in the 
Ni-C$_{60}$-Ni system \cite{fdotf_exp} in the Kondo regime.

\begin{figure}[b!]
\resizebox{0.45\textwidth}{!}{\includegraphics{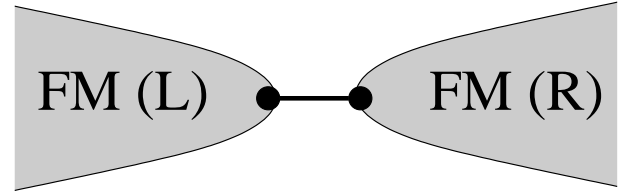}}
\caption{Schematic display of nanoscale magnetic tunnel junction 
\label{model}}
\end{figure}

 In this paper we propose theoretically another possible mechanism for the inverse TMR, based
on the simple model study. 
The model system is schematically displayed in Fig.~\ref{model}, where the spin polarized
electrons of one FM electrode pass through a narrow channel into the second FM electrode.
When the transmission probability is small between two FM electrodes, 
the TMR is normal or positive. 
On the other hand the sign of TMR can become opposite to the case of weak tunneling 
when the transmission probability is large close to a unity.

 The rest of this paper is organized as follows. In Sec. II, the model Hamiltonian 
is introduced for the nanoscale magnetic tunnel junctions
and the spin-polarized current is formulated using the Keldysh nonequilibrium Green's 
function technique. The results of our work are presented in Sec. III and 
a conclusion is included in Sec. IV.  In Appendix, we present the tight-binding Hamiltonian 
approach to the inverse TMR effect in the nanoscale magnetic tunnel junctions.

\section{Formalism}   
 The considered model system is the magnetic tunnel junction with a very small 
junction area (linear dimension is of the order of a few Fermi wavelengths).
When the junction area is very large compared to the Fermi wave length scale, 
the Slonczewski model \cite{slonczewski} will be relevant. 
The ferromagnetic (FM) metals, which we conveniently call the left ($p=L$) 
and right ($p=R$) electrodes, are described by the two conduction bands of majority 
and minority spins. 
The relative direction of magnetizations in two electrodes is chosen to be arbitrary. 
It is convenient to write the spin-polarized conduction band in the diagonal basis 
or the spin quantization direction is chosen to be the direction of magnetization.
\begin{eqnarray}
\label{band}
H_p &=& \sum_{k\sigma} \epsilon_{pk\sigma} c_{pk\sigma}^{\dag} c_{pk\sigma}.
\end{eqnarray} 
Here $c_{pk\sigma}^{\dag}$ and $c_{pk\sigma}$ are the creation and annihilation operators, 
respectively, for electrons of wave number $k$ in the electrode $p=L,R$, with the spin direction 
$\sigma = \pm$. $\sigma = +(-)$ means the spin is aligned parallel (antiparallel) to 
the direction of magnetization.
$\epsilon_{pk\sigma}$ is the energy dispersion relation
for electrons in the ferromagnetic metals.

 Since the magnetization directions in the left and right electrodes or the spin quantization
axes are different, we must be careful when writing the coupling Hamiltonian 
between two FM electrodes.  
The flow of electrons from one FM lead to the other is modeled by the transfer Hamiltonian.
\cite{transfer_ham1, transfer_ham2}
\begin{eqnarray}
\label{transfer}
H_1 &=& \sum_{kk'}\sum_{\alpha\beta} \left[ 
     t_{Lk\alpha,Rk'\beta} c_{Lk\alpha}^{\dag} c_{Rk'\beta} + H.c. \right].
\end{eqnarray}
The spin quantization axis (the positive $z$ axis) in this transfer Hamiltonian is chosen 
to be the direction of the current flow from left to right. In Eq.~(\ref{transfer}), 
the spin indices $\alpha, \beta = \uparrow, \downarrow$. 
$\alpha = \uparrow (\downarrow)$ means the spin is directed parallel (antiparallel) to 
the positive $z$ direction. Note that we are completely free to choose any direction 
as the spin quantization direction for this transfer Hamiltonian. 
After fixing the positive $z$ axis, we can select the orthogonal $x$ and $y$ axes, 
perpendicular to the $z$ axis, at our disposal. 
Once the coordinate system is fixed, the direction of magnetization in each FM electrode 
can be specified by two angles $(\theta_p, \phi_p)$ in the spherical coordinates
with $p=L,R$ (left, right electrode). Here $\theta_p$ measures the angle between the magnetization
direction and our chosen $z$ axis, and $\phi_p$ is the azimuthal angle of the magnetization 
measured from the $x$ axis. 
$t_{Lk\alpha,Rk'\beta}$ in Eq.~(\ref{transfer}) is the hopping integral from the spin band $\beta$ 
in the right electrode to the spin band $\alpha$ in the left. 
At this point we consider the most general form of the hopping integral which allows the spin flips. 
The first term in Eq.~(\ref{transfer}) represents the transfer of electrons 
from the right electrode ($R$) with the spin direction $\beta$ 
to the left ($L$) lead with the spin direction $\alpha$.
This transfer Hamiltonian is relevant when the transfer of electrons from one lead 
to the other occurs dominantly through a very narrow channel (one transport channel).

 It is straightforward to show that the electron annihilation operators 
in two different bases of Eqs.~(\ref{band}) and (\ref{transfer}) are related to each other 
by the equation
\begin{eqnarray}
\label{utr}
\begin{pmatrix} c_{pk+} \cr c_{pk-} \end{pmatrix}
  &=& \begin{pmatrix} \cos\frac{\theta_p}{2}  &  e^{-i\phi_p} \sin\frac{\theta_p}{2}  \cr
        - e^{i\phi_p} \sin\frac{\theta_p}{2}  &  \cos\frac{\theta_p}{2}
      \end{pmatrix}
   \begin{pmatrix} c_{pk\uparrow} \cr c_{pk\downarrow} \end{pmatrix}.
\end{eqnarray}  
Let us introduce the electron spinors ($\Phi$ and $\Psi$) and the unitary transformation 
matrix ($U$).
\begin{subequations}
\begin{eqnarray}
\Phi_{pk} &\equiv& \begin{pmatrix} c_{pk+} \cr c_{pk-} \end{pmatrix}, ~~~
  \Psi_{pk} ~\equiv~ \begin{pmatrix} c_{pk\uparrow} \cr c_{pk\downarrow} \end{pmatrix}, \\
U_p &\equiv& \begin{pmatrix} \cos \frac{\theta_p}{2} & 
                   - e^{-i\phi_p} \sin\frac{\theta_p}{2} \cr
             e^{i\phi_p} \sin\frac{\theta_p}{2} & \cos \frac{\theta_p}{2}
          \end{pmatrix}.
\end{eqnarray}
\end{subequations}
The electrode Hamiltonian, Eq~(\ref{band}), is diagonal in the $\Phi$ basis, but not diagonal
in the $\Psi$ basis. The two bases are related by the equation, $\Psi_{pk} = U_p \Phi_{pk}$, 
which is none other than Eq.~(\ref{utr}). Using the spinor operators, the model Hamiltonian
can be written as
\begin{subequations}
\begin{eqnarray}
H_p &=& \sum_k \Phi_{pk}^{\dag} \begin{pmatrix} \epsilon_{pk+} & 0 \cr 0 & \epsilon_{pk-}
            \end{pmatrix} \Phi_{pk},  \\ 
H_1 &=& \sum_{kk'} \left[ \Psi_{Lk}^{\dag} V_{Lk,Rk'} \Psi_{Rk'} + H.c. \right],  \\
%                         + \Psi_{Rk'}^{\dag} V_{Rk',Lk} \Psi_{Lk} \right], \\
V_{Lk,Rk'} &=& \begin{pmatrix} t_{Lk\uparrow,Rk'\uparrow} & t_{Lk\uparrow,Rk'\downarrow} \cr
                t_{Lk\downarrow,Rk'\uparrow} & t_{Lk\downarrow,Rk'\downarrow}
           \end{pmatrix}.
\end{eqnarray}
\end{subequations}
Here we introduced the hopping matrix $V_{Lk,Rk'}$.

 We are now in a position to derive the expression of spin-polarized current for our model system.
Using the nonequilibrium Green's function method,\cite{neqgreen,neqgreen1} 
the electric current 
flowing from left to right can be computed by thermally averaging the current operator
 \cite{neqgreen2,neqgreen3}
\begin{subequations}
\begin{eqnarray}
\hat{I} &=& \frac{e}{i\hbar} \sum_{kk'}\sum_{\alpha\beta} \left[ 
     t_{Lk\alpha,Rk'\beta} c_{Lk\alpha}^{\dag} c_{Rk'\beta} 
    - H.c. \right]. 
\end{eqnarray}
It is now more convenient to write the current operator in the matrix form. 
\begin{eqnarray}
\hat{I} &=& \frac{e}{i\hbar} \sum_{kk'} \left[ \Psi_{Lk}^{\dag} V_{Lk,Rk'} \Psi_{Rk'} 
                         - \Psi_{Rk'}^{\dag} V_{Rk',Lk} \Psi_{Lk} \right].
\end{eqnarray}
\end{subequations}
Let us now introduce the time-ordered mixed Green's functions\cite{neqgreen,neqgreen1} 
which are defined by 
\begin{subequations}
\begin{eqnarray}
i\hbar G_{LR} (kt, k't') 
  &=& \langle T \Psi_{Lk} (t) \Psi_{Rk'}^{\dag} (t') \rangle,  \\
i\hbar G_{RL} (kt, k't') 
  &=& \langle T \Psi_{Rk} (t) \Psi_{Lk'}^{\dag} (t') \rangle,
\label{mixGRL}
\end{eqnarray}
\end{subequations}
where the symbol $\langle A \rangle$ means the thermal average of $A$ and $T$ 
inside the thermal average means the time-ordering in the Keldysh contour. 
In terms of the above mixed Green's functions, the current can be written as
\begin{eqnarray}
\label{current}
I &=& 2e \sum_{kk'} \mbox{Im} \mbox{Tr} \left[ V_{Lk,Rk'} G_{RL}^{<} (k't, kt) \right]
    \nonumber\\
 &=& \frac{2e}{h} \int d\epsilon \sum_{kk'} \mbox{Im} \mbox{Tr} \left[ V_{Lk,Rk'} 
       G_{RL}^{<} (k', k; \epsilon) \right].
\end{eqnarray}
The second line is obtained from the first line after the Fourier transform from time to energy
variables.

 Since we are mainly interested in the linear response conductance of the system, the physical 
properties are determined by the energy structure near the Fermi level at low temperature
compared with the Fermi temperature. This is the case we are going to consider below. 
Since the transport properties are mainly determined by electrons near the Fermi level,
the dependence of the hopping integral on the wave vector may be neglected and 
the hopping integral can be replaced by its value at the Fermi energy. 
From now on, the dependence of the hopping matrix $V_{LR}$ on the wave vector will be 
suppressed under this approximation scheme. In the tight-binding Hamiltonian approach, 
the hopping matrix $V_{LR}$ does not depend on the wave vector (see the Appendix for details.) 
Under this approximation, it is convenient to introduce the wave-vector-summed Green's functions
\cite{neqgreen3} to simplify the algebra.  
\begin{subequations}
\begin{eqnarray}
G_p (t,t') &=& \sum_{k} G_p(k; t,t'), ~~~ p = L, R, \\
i\hbar G_p(k; t,t')
  &\equiv& \langle T \Psi_{pk} (t) \Psi_{pk}^{\dag} (t') \rangle,  \\
G_{RL} (t,t') &=& \sum_{kk'} G_{RL} (kt; k't').
\end{eqnarray}
\end{subequations}
Note that $G_p(k; t,t')$ is the Green's function of electrons in each lead $p=L,R$ when
the coupling between two FM electrodes is absent, 
while $G_{RL} (kt; k't')$ is defined in Eq.~(\ref{mixGRL}).
Using the Feynman diagrams, we can readily show that the mixed Green's function is determined 
by the Dyson-like equation \cite{neqgreen3}
\begin{widetext}
\begin{eqnarray}
G_{RL} (t,t') 
 &=& G_{RL0} (t,t') + \int_C dt_1 G_{RL0} (t,t_1) V_{LR} G_{RL} (t_1, t'), 
\end{eqnarray}
%
%
%\end{widetext}
%
%
where the auxiliary Green's function $G_{LR0}$ is defined as $G_{RL0} (t,t')
 = \int_C dt_1 G_R(t,t_1) V_{RL} G_L(t_1, t')$. 
Here $C$ in the integral sign is the Keldysh contour embracing the real time axis.
The Green's functions $G_{L/R}$ in the above equations can be interpreted as the Green's function 
of an electron located at the local tip site of the left/right lead in the absence of 
the coupling between two FM leads (see the Appendix.) 
After analytically continuing \cite{langreth} to the real time axis from the Keldysh contour, 
it is straightforward to find the lesser Green's function $G_{RL}^{<}$
\begin{eqnarray}
\label{GRLless}
G_{RL}^{<} (\epsilon) 
 &=& \left[ 1 - G_{RL0}^{r} (\epsilon) V_{LR} \right]^{-1} G_{RL0}^{<} (\epsilon)  %\nonumber\\
% && \times  
 \left[ 1 - V_{LR} G_{RL0}^{a} (\epsilon) \right]^{-1}, 
\end{eqnarray}
where the axillary Green's functions are 
\begin{subequations}
\label{GRL0}
\begin{eqnarray}
G_{RL0}^{r,a} (\epsilon) &=& G_{R}^{r,a} (\epsilon) V_{RL} G_{L}^{r,a} (\epsilon),  \\
G_{RL0}^{<} (\epsilon) &=& G_R^{<}(\epsilon) V_{RL} G_L^{a}(\epsilon) 
   + G_R^{r}(\epsilon) V_{RL} G_L^{<}(\epsilon).
\end{eqnarray}
\end{subequations}
The $G^{<, r, a}$ denote the lesser, retarded, and advanced Green's functions, respectively.

 To find the expression of the spin-polarized current, we need the explicit forms of 
the lesser, retarded, and advanced Green's functions for electrons in the FM electrode.
We have to note that the Green's functions of the FM electrodes in the above equations
are not diagonal, because we worked in the spin basis of the transfer Hamiltonian. 
\begin{subequations}
\begin{eqnarray}
i\hbar G_p (k; t,t') 
  &=& \langle T \Psi_{pk} (t) \Psi_{pk}^{\dag} (t') \rangle   %\nonumber\\
  ~=~ U_p \langle T \Phi_{pk} (t) \Phi_{pk}^{\dag} (t') \rangle U_p^{\dag},  \\
i\hbar g_p (k; t,t') 
  &=& \langle T \Phi_{pk} (t) \Phi_{pk}^{\dag} (t') \rangle
  ~=~ \begin{pmatrix} \langle T c_{pk+} (t) c_{pk+}^{\dag} (t') \rangle  & 0 \cr
                     0 & \langle T c_{pk-} (t) c_{pk-}^{\dag} (t') \rangle
      \end{pmatrix}.
\end{eqnarray}  
\end{subequations}
Here $g_p$ is the Green's functions of the FM electrodes in the spin diagonal basis and related 
to $G_p$ by the unitary transformation, $G_p = U_p g_p U_p^{\dag}$. 
Using the above relations, the desired Green's functions can be found for the model Hamiltonian 
of Eq.~(\ref{band})
%
%
%\begin{widetext}
%
%
\begin{subequations}
\begin{eqnarray}
\label{Gbandra}
G_p^{r,a} (\epsilon)
  &=& U_p \sum_k \begin{pmatrix} 
     [\epsilon - \epsilon_{pk+} \pm i \delta]^{-1}  & 0 \cr
      0 &  [\epsilon - \epsilon_{pk-} \pm i\delta]^{-1}
    \end{pmatrix} U_p^{\dag},  \\
\label{Gbandless}
G_p^{<} (\epsilon) 
  &=& 2\pi f_p (\epsilon) U_p \sum_{k} \begin{pmatrix} 
      \delta(\epsilon - \epsilon_{pk+})  & 0 \cr  0 &  \delta(\epsilon - \epsilon_{pk-})
     \end{pmatrix} U_p^{\dag}.
\end{eqnarray}
\end{subequations}
\end{widetext}
Here an infinitesimally small positive number $\delta$ is included for the definition 
of the retarded ($+$ sign) and advanced ($-$ sign) Green's functions, $G_p^{r,a}$. 
$f_p (\epsilon) = f(\epsilon-\mu_p)$ is the Fermi-Dirac thermal distribution 
function in the lead $p=L,R$ and $\mu_p$ is the chemical potential shift of each lead
caused by the source-drain bias voltage.
Let us define $R_{p\pm}$ and $N_{p\pm}$ as
\begin{subequations}
\begin{eqnarray}
R_{p\pm} (\epsilon) &=& {\cal P} \sum_k \frac{1}{\epsilon - \epsilon_{pk\pm}},  \\
N_{p\pm} (\epsilon) &=& \sum_k \delta (\epsilon - \epsilon_{pk\pm}),
\end{eqnarray}
\end{subequations}
where ${\cal P}$ means the principal part of the integral. $N_{p\pm} (\epsilon)$ 
is none other than the density of states (DOS) for the majority/minority spin 
in the FM lead $p=L,R$.
The retarded, advanced, and lesser Green's functions [Eqs.~(\ref{Gbandra}) and (\ref{Gbandless})] 
can be written as 
\begin{subequations}
\begin{eqnarray}
G_p^{r,a} (\epsilon)
  &=& U_p [ R_p \mp i \pi N_p ] U_p^{\dag},  \\
\label{Gbandless1}
G_p^{<} (\epsilon) 
  &=& 2\pi f_p (\epsilon) U_p N_p U_p^{\dag},
\end{eqnarray}
\end{subequations}
 where $R_p = \left( \begin{smallmatrix} R_{p+} & 0 \cr 0 & R_{p-} \end{smallmatrix} \right)$ 
and $N_p = \left( \begin{smallmatrix} N_{p+} & 0 \cr 0 & N_{p-} \end{smallmatrix} \right)$.
When the Fermi level lies inside the band, $R_{p\pm} (\epsilon=E_F)$ is expected to be small due to 
the mutual cancellation of electron (above the Fermi level) and hole (below the Fermi level) 
contributions. 
In this case or in a wide conduction band limit,\cite{wideband}
 we can set $R_{p\pm} (\epsilon=E_F) \approx 0$.
The retarded and advanced Green's functions are given in simple forms in this case. 
\begin{eqnarray}
\label{Gbandra_sim}
G_p^{r,a} (\epsilon) 
  &=& \mp i \pi U_p N_p U_p^{\dag}.
\end{eqnarray}
For half metals with only one spin band occupied at the Fermi level, 
the value of $R$ for unoccupied spin band may not be neglected 
such that the retarded and advanced Green's functions may have real component.

 When the Fermi level lies well inside the band for both spin directions or 
in a wide band limit,\cite{wideband} we can approximately set $R_{p\pm} \approx 0$.  
In this case, inserting the Eqs.~(\ref{Gbandless1}) and (\ref{Gbandra_sim}) into Eqs.~(\ref{GRL0}),
the expressions of $G_{RL0}^{r,a}$ and $G_{RL0}^{<}$ can be simplified as
\begin{subequations}
\label{GRL01}
\begin{eqnarray}
G_{RL0}^{r,a} (\epsilon)
 &=& - \pi^2 \hat{N}_R V_{RL} \hat{N}_L, \\
G_{RL0}^{<} (\epsilon)
 &=& 2i\pi^2 \left[ f_R(\epsilon) - f_L(\epsilon) \right] \hat{N}_R V_{RL} \hat{N}_L, 
\end{eqnarray}
\end{subequations}
where $\hat{N}_p = U_p N_p U_p^{\dag}$. 
Inserting these two Eqs.~(\ref{GRL01}) into Eq.~(\ref{GRLless}), we find for the lesser
Green's function $G_{RL}^{<}$
\begin{eqnarray}
G_{RL}^{<} &=& 2i\pi^2 \left[ f_R(\epsilon) - f_L(\epsilon) \right] 
  \left[ 1 + \pi^2 \hat{N}_R V_{RL} \hat{N}_L V_{LR} \right]^{-1}  \nonumber\\
 && \times  \hat{N}_R V_{RL} \hat{N}_L 
  \left[ 1 + \pi^2 V_{LR} \hat{N}_R V_{RL} \hat{N}_L  \right]^{-1}.
\end{eqnarray}
Combining this lesser Green's function and Eq.~(\ref{current}), the expression of the spin-polarized
current can be written as
\begin{eqnarray}
I &=& \frac{2e}{h} \int d\epsilon \mbox{Im} \mbox{Tr} \left[ V_{LR} 
     G_{RL}^{<} (\epsilon) \right]   \nonumber\\
 &=& \frac{2e}{h} \int d\epsilon \left[ f_R (\epsilon) - f_L (\epsilon) \right]
       T (\epsilon), 
\end{eqnarray}
where 
\begin{subequations}
\begin{eqnarray}
\label{tprob}
T (\epsilon) &=& 2 \mbox{Re}\mbox{Tr} [1 + \Gamma]^{-1} \Gamma [1 + \Gamma]^{-1},  \\
\Gamma &=& \pi^2 \hat{N}_R V_{RL} \hat{N}_L V_{LR}.
\end{eqnarray}
\end{subequations}
In obtaining the above form of $T(\epsilon)$, the trace property 
of the matrix product was invoked. 
Here $T (\epsilon)$  is the transmission probability  through a narrow channel
for spin-polarized electrons, when the magnetization directions of two FM electrodes are
arbitrary. 
The information about the orientation of two magnetizations is included in the DOS matrices,
$\hat{N}_L$ and $\hat{N}_R$. 
The matrix $\Gamma$ measures the effective transfer rate of spin-polarized electrons 
from one FM electrode to the other FM electrode, and contains the dependence on 
the relative orientation of two magnetizations. 
The above form of the transmission probability is valid 
for the wide band case or when the Fermi level lies well inside the bands 
for both spin directions.
The linear response conductance $G$ is then given by the expression,
$G = \frac{2e^2}{h} \cdot T(\epsilon=0)$.
Below we are going to confine our interest to the linear response regime and to the case
of a wide band limit.

  We can give some physical meaning to the density matrix $\hat{N}_p = U_p N_p U_p^{\dag}$.
After some algebra, $\hat{N}_p$ can be expressed as 
\begin{subequations}
\begin{eqnarray}
\hat{N}_p &=& N_{p0} [ 1 + P_p \hat{n}_p \cdot \vec{\sigma} ], \\
N_{p0} &=& = \frac{1}{2} [ N_{p+} + N_{p-} ], \\
P_p &=& \frac{ N_{p+} - N_{p-} }{  N_{p+} + N_{p-} }.
\end{eqnarray}
\end{subequations}
Here $\vec{\sigma}$ is the Pauli matrices, $N_{p0}$ is the average DOS
of the majority and minority spin bands. 
$P_p$ measures the spin polarization at a given energy $\epsilon$ of the FM electrode
$p=L,R$ and agrees with the definition of the spin polarization in the Julliere model. 
\cite{julliere} 
$\hat{n}_p$ is the unit vector defined by the two angles, $\theta_p$ and $\phi_p$
in the spherical coordinate system and is aligned parallel to the magnetization direction 
of the $p=L,R$ FM electrode.

\section{Results and discussion}
 In this section we study the TMR behavior for the wide band case with the transfer Hamiltonian:
$t_{L\alpha, R\beta} = t_{LR} \delta_{\alpha\beta}$. 
 Electrons do not flip their spin while passing through the narrow channel.
In this case the expression of $\Gamma$ in the transmission probability, Eq.~(\ref{tprob}), 
is simplified as
\begin{eqnarray}
\Gamma &=& \pi^2 |t_{LR}|^2 \hat{N}_R \hat{N}_L   \nonumber\\
   &=& \gamma (1 + P_R \hat{n}_R \cdot \vec{\sigma}) 
      (1 + P_L \hat{n}_L \cdot \vec{\sigma}). 
\end{eqnarray}
Here $\gamma \equiv \pi^2 N_{L0} N_{R0} |t_{LR}|^2$ measures the average transfer rate of electrons
for two spin directions.
The expression of the transmission probability $T$ is 
\begin{widetext}
\begin{eqnarray}
\label{gencond}
T(\epsilon) &=& 4 \gamma \times \frac{1 + \vec{P}_L \cdot \vec{P}_R 
      + \gamma (1 - P_L^2)(1 - P_R^2) (2 + \gamma + \gamma \vec{P}_L \cdot \vec{P}_R) }
    { [ 1 + 2 \gamma (1 + \vec{P}_L \cdot \vec{P}_R) 
        + \gamma^2 (1 - P_L^2)(1 - P_R^2) ]^2 }.
\end{eqnarray}
\end{widetext}
Note that $\vec{P}_L = \hat{n}_L P_L$ and $\vec{P}_R = \hat{n}_R P_R$.
This is the most general result for the magnetic tunnel junctions with a narrow 
junction area. Note that the conductance, $2e^2/h \times T(0)$, depends only 
on the relative angle of two magnetization directions. 
In the weak tunneling limit or $\gamma \ll 1$, the transmission probability 
can be approximated as
$T = 4 \gamma ( 1 + \vec{P}_L \cdot \vec{P}_R ) = 4 \gamma ( 1 + P_L P_R \cos\theta)$.
The conductance is then given by the expression, 
$G = G_0 ( 1 + P_L P_R \cos\theta)$ with $G_0 = 8\gamma e^2/h$.
This is the familiar expression of the conductance which is in agreement with 
the Julliere model.\cite{julliere} The TMR ratio is then $TMR = 2P_L P_R/(1-P_LP_R)$.
In the weak tunneling limit, the TMR is determined solely by the polarization of
the ferromagnetic metals. In the strong tunneling limit the angle dependence 
of the conductance is still an even function of $\theta$, but is more complicated
as shown in Eq.~(\ref{gencond}).

 When two ferromagnetic electrodes are the same 
($N_{L+} = N_{R+} = N_{+}$ and $N_{L-} = N_{R-} = N_{-}$) with the possible difference
in the magnetization direction, we have $P_L = P_R = P$. 
The expression of $T$, Eq.~(\ref{gencond}), is further simplified
\begin{widetext}
\begin{eqnarray}
\label{samecond}
T(\cos\theta) &=& 4 \times \frac{ \gamma (1 + P^2 \cos\theta)
      + \gamma^2 (1 - P^2)^2 (2 + \gamma [ 1 +  P^2 \cos\theta]) }
   {\left[ 1 + 2\gamma (1 + P^2\cos\theta) + \gamma^2(1 - P^2)^2 \right]^2}.
\end{eqnarray}
\end{widetext}
Especially for the parallel ($\cos\theta = 1$) and antiparallel ($\cos\theta = -1$) configurations, 
the transmission probabilities, respectively, given by the expressions
\begin{subequations}
\begin{eqnarray}
\label{T_P}
T_P &=& \frac{2 \gamma_{+}} { ( 1 + \gamma_{+})^2} 
   + \frac{2 \gamma_{-}} { ( 1 + \gamma_{-})^2}, \\
T_{AP} &=& \frac{4 \sqrt{\gamma_{+}\gamma_{-}} } { ( 1 + \sqrt{\gamma_{+}\gamma_{-}} )^2}. 
\label{T_AP}
\end{eqnarray}
\end{subequations}
Here $\gamma_{\pm}$ is defined by the equation
$\gamma_{\pm} \equiv \pi^2 N_{\pm}^2 |t_{LR}|^2$ ($0 < \gamma_{\pm} \leq 1$), 
and is related to the transmission probability $T_{\pm}$ by the relation
\begin{eqnarray}
T_{\pm} &=& \frac{4\gamma_{\pm} }{ ( 1 + \gamma_{\pm} )^2 }. 
\end{eqnarray}
$T_{+}$ is the transmission probability for electrons from the majority band in the left electrode
to the majority band in the right electrode, while $T_{-}$ from minority to minority. 
Note that the average transfer rate $\gamma$ and the polarization $P$ are related to 
the transfer rates of majority and minority spins by the equations
\begin{subequations}
\begin{eqnarray}
\gamma &=& \frac{1}{4} [ \sqrt{\gamma_{+}} + \sqrt{\gamma_{-}} ]^2,   \\
P &=& \frac{ \sqrt{\gamma_{+}} - \sqrt{\gamma_{-}} }{ \sqrt{\gamma_{+}} + \sqrt{\gamma_{-}} }.
\end{eqnarray}
\end{subequations}
The transmission probabilities, $T_P$ and $T_{AP}$, can be expressed in terms of 
$T_{\pm}$ as
$T_P = \frac{1}{2} ( T_{+} + T_{-} )$ and
$T_{AP} = 4\sqrt{T_{+} T_{-}} ( 1 + \sqrt{1-T_{+}} ) ( 1+ \sqrt{1-T_{-}} )
  [ \sqrt{T_{+}T_{-}} + ( 1 + \sqrt{1-T_{+}})(1+ \sqrt{1-T_{-}}) ]^{-2}$.
%Below we analyze the TMR in detail.

{\it In the weak tunneling limit} $T_{\pm} \ll 1$ or $\gamma_{\pm} \ll 1$, 
the Eq.~(\ref{samecond}) can be approximated as 
$T (\cos \theta) = 4\gamma ( 1+ P^2 \cos\theta)$.
Especially for the parallel and antiparallel configurations, we have
$T_P = 2 (\gamma_{+} + \gamma_{-})$ and 
 $T_{AP} = 4\sqrt{\gamma_{+} \gamma_{-}}$.
Since $\gamma_{+} + \gamma_{-} \geq 2\sqrt{\gamma_{+} \gamma_{-}}$, 
we have the conductance inequality $G_{P} \geq G_{AP}$. 
This inequality relation is true generally when the transmission probability is weak.
That is, the current flows more in the parallel configuration than in the antiparallel 
configuration.

\begin{figure}[t!]
\includegraphics{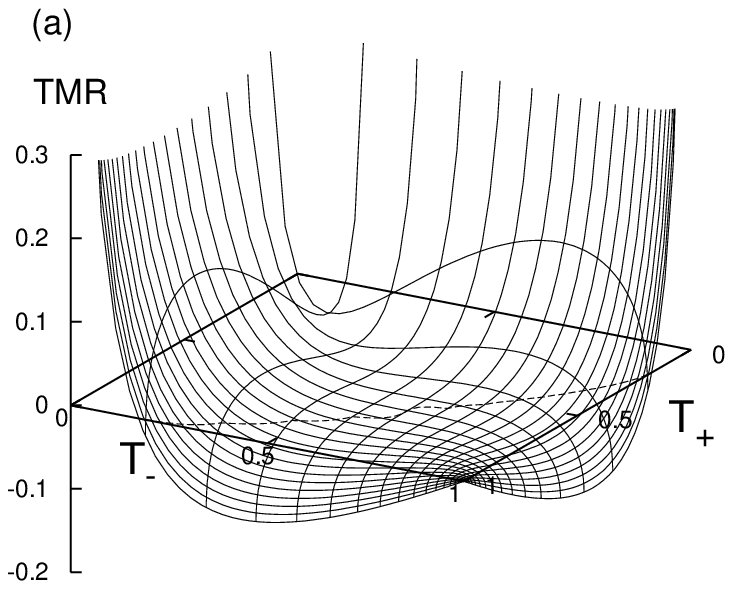} 
\vskip -1.0cm 
\resizebox{0.4\textwidth}{!}{\includegraphics{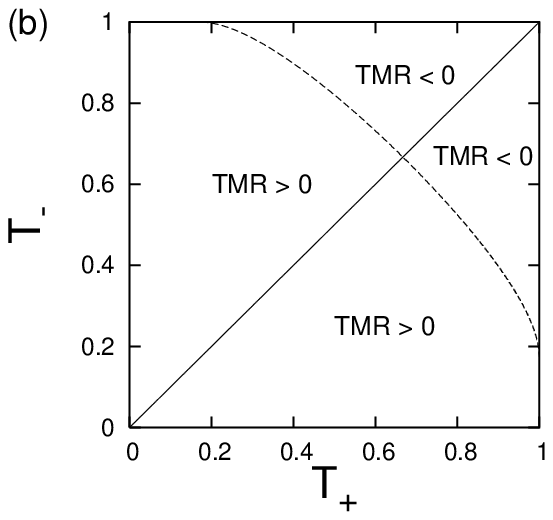}} 
\caption{Inverse TMR. (a) Dependence of TMR on the transmission probabilities of electrons 
 for majority ($T_{+}$) and minority spin ($T_{-}$). 
 The dashed line is the boundary between the positive and negative TMR. 
(b) Sign of TMR in parameter space of $(T_{+}, T_{-})$. 
TMR vanishes along the solid and dashed lines.
Two lines meet at $T_{+} = T_{-} = 2/3$. 
 The TMR becomes negative when the tunneling probabilities for electrons of both spin directions
 is close to a unity. The curve $T_{+} = T_{-}$ (solid line) corresponds to the nonmagnetic 
 electrodes.
\label{tmr}}
\end{figure}

{\it In the strong tunneling limit}, the above inequality in conductance can be 
reversed and more current can flow in the antiparallel alignment than in the parallel
alignment. To see when the current flows more in the AP than in the P, we study 
$T_P$ and $T_{AP}$ [Eqs.~(\ref{T_P}) and (\ref{T_AP})] in the plane of 
$(\gamma_{+}, \gamma_{-})$ or $(T_{+}, T_{-})$. 
The curves on which $T_P = T_{AP}$ are given by the equations
\begin{subequations}
\begin{eqnarray}
0 &=& \gamma_{+} - \gamma_{-},  \\
\label{2ndcurve}
0 &=& (\gamma_{+}\gamma_{-} - 1)^2 
    - 2\sqrt{\gamma_{+}\gamma_{-}} (1+\gamma_{+})(1+\gamma_{-}).
\end{eqnarray}
\end{subequations}
Along these two curves the conductance in the P and AP configurations is equal. 
The curve $\gamma_{+}=\gamma_{-}$ corresponds to the case that $T_P = T_{AP}$ and 
the magnetic polarization at the Fermi energy is zero or $P=0$.
 The second curve, Eq.~(\ref{2ndcurve}), is the boundary between two regions:
$T_P > T_{AP}$ (close to the origin) and $T_P < T_{AP}$ (strong tunneling regime),
and corresponds to the case that $P \neq 0$, but $T_P = T_{AP}$.
Two curves meet each other at 
$(\gamma_{+}, \gamma_{-}) = (2-\sqrt{3}, 2-\sqrt{3})$, 
which corresponds to the point $(T_{+}, T_{-}) = (2/3, 2/3)$ 
in the plane of transmission probabilities. 
The behavior of TMR is summarized in Fig.~\ref{tmr}(a) and (b).
The value of TMR is plotted as a function of $(T_{+}, T_{-})$ in the panel (a) and 
the boundary between normal and inverse TMR is marked by the dashed line. 
The panel (b) provides the TMR diagram in the $(T_{+}, T_{-})$ plane. 
Especially in the strong tunneling limit, the inverse TMR can be realized 
in the nanoscale magnetic tunnel junctions.

\begin{figure}[t!]
\includegraphics{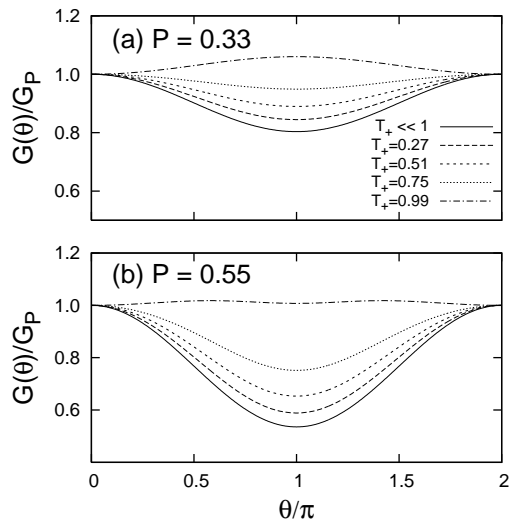} 
\caption{Dependence of conductance on the relative angle between two magnetization directions
with varying transmission probability $T_{+}$ for the majority spin.
In panel (a) $P=33\%$ is chosen to simulate the Ni electrodes,\cite{review} while
$P=55\%$ in panel (b) for CoFe electrodes. The solid line ($T_{+} << 1$) corresponds to 
the Julliere result or the very weak tunneling limit.
\label{gratio}}
\end{figure}

 Fig.~\ref{gratio} displays the conductance ratio, $G(\theta)/G_P$, 
as a function of the relative angle between two directions of magnetization
[$G_P = G(\theta=0)]$. While the polarization is fixed, the transmission probabilities
are varied. In the weak
tunneling limit the solid curve traces the cosine function. 
With increasing transmission probabilities, the curves deviate from the Julliere
result and furthermore the conductance minimum can change into a maximum 
leading to the inverse TMR.

\section{Summary and conclusion}
 To summarize, we studied the spin-polarized transport through a narrow channel using 
the transfer Hamiltonian approach and the nonequilibrium Green's function method. 
Our analysis applies to the case when the Fermi level lies well inside both majority and minority 
spin bands or to the wide band limit. 
In the weak tunneling limit we reproduce the Julliere results and the TMR is positive.
On the other hand, in the strong tunneling limit 
the inverse TMR opposite to the weak tunneling case is found. 
We presented another theoretical scenario for the inverse TMR effect. 
Our model study may be relevant to the magnetic junctions with a
quantum point contact between two ferromagnetic metals or to the thick planar MTJs with a pinhole
which accommodate only one transport channel.

\acknowledgments
This work was supported by Korea Research Foundation Grant (KRF-2003-042-C00038), 
grant No. R01-2005-000-10303-0 from the Basic Research Program of the Korea Science 
\& Engineering Foundation, and KIST Vision21 program.

\appendix

\begin{widetext}

\section{Tight-binding model for the ferromagnetic electrodes}
 To illustrate the relevance of our approach to real systems, we consider the description of 
the ferromagnetic (FM) electrodes in terms of the tight-binding Hamiltonian. 
Two FM electrodes are coupled by the hopping term between two atomic sites. 
Though our formalism can be generally applied to any types of lattice, we confine our study 
to the lattices for which the analytic solution of the Green's functions can be readily found. 
The examples include the Bethe lattice out of which the simplest one is the semi-infinite 
one-dimensional tight-binding atomic chain (see Fig.~\ref{chain}.)

\begin{figure}[b!]
\resizebox{0.45\textwidth}{!}{\includegraphics{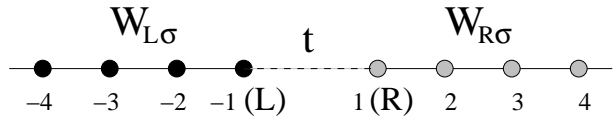}}
\caption{Tight-binding model for the FM electrodes. The FM electrodes are described 
by the semi-infinite chain. The site of $i = -1 (i = 1)$ is the terminal site of the left (right) 
electrode, and is denoted as $i=L (i=R)$ for the notational convenience. 
\label{chain}}
\end{figure}

 The left and right electrodes are described by the tight-binding Hamiltonian
\begin{subequations}
\label{band_A}
\begin{eqnarray}
H_L &=& \sum_{i \leq -1} \sum_{\sigma=\pm} \epsilon_{L\sigma} c_{i\sigma}^{\dag} c_{i\sigma} 
   - \sum_{i \leq -2} \sum_{\sigma = \pm} 
      \left[ W_{L\sigma} c_{i+1\sigma}^{\dag} c_{i\sigma} + H.c. \right],   \\
H_R &=&  \sum_{i \geq 1} \sum_{\sigma=\pm} \epsilon_{R\sigma} c_{i\sigma}^{\dag} c_{i\sigma} 
   - \sum_{i \geq 1} \sum_{\sigma = \pm} 
      \left[ W_{R\sigma} c_{i+1\sigma}^{\dag} c_{i\sigma} + H.c. \right].
\end{eqnarray}
\end{subequations}
The left (right) FM electrode is terminated at the site of $i = -1$ or $i = L$ ($i = 1$ or $i = R$),
respectively. The $i=L$ site will be called the left terminal site and the $i=R$ site 
the right terminal site. 
The electron creation ($c_{i\sigma}^{\dag}$) and annihilation ($c_{i\sigma}$) 
operators are defined in the spin diagonal basis. 
$\sigma = + (-)$ means the spin is aligned parallel (antiparallel) to the magnetization, 
respectively. However the spin quantization directions
of two leads are different due to different magnetization directions. 
The on-site energies, $\epsilon_{L\sigma}$ and $\epsilon_{R\sigma}$, include 
the exchange splitting between two spin directions as well as a possible voltage
difference between two FM leads.
In order to take into account the different bandwidth of spin-up and spin-down bands, 
the hopping integrals, \cite{spinband}
 $W_{L\sigma}$ and $W_{R\sigma}$, include the spin index $\sigma = \pm$.
Magnetic polarization is determined by the values of on-site energies and hopping integrals.

 Adopting the same convention for the direction of spin quantization
(chosen to be the direction of the current flow from left to right) 
as in the main text, the hopping term which is responsible for the flow of electrons 
between two leads can be written as
\begin{eqnarray}
\label{transfer_A}
H_1 &=& \sum_{\alpha\beta} \left[ 
     t_{L\alpha,R\beta} c_{L\alpha}^{\dag} c_{R\beta} + H.c. \right].
\end{eqnarray}
Here $\alpha, \beta = \uparrow, \downarrow$ and the $\uparrow (\downarrow)$ direction is 
parallel (antiparallel) to the spin quantization direction. 
Since the spin quantization directions in Eqs.~(\ref{band_A}) and Eq.~(\ref{transfer_A}) are 
different, the electron operators in two spin bases are related to each other 
by the unitary transformation. For details, see the main text.
\begin{subequations}
\begin{eqnarray}
\Phi_{p} &\equiv& \begin{pmatrix} c_{p+} \cr c_{p-} \end{pmatrix}, ~~~
  \Psi_{p} ~\equiv~ \begin{pmatrix} c_{p\uparrow} \cr c_{p\downarrow} \end{pmatrix}, \\
U_p &\equiv& \begin{pmatrix} \cos \frac{\theta_p}{2} & 
                   - e^{-i\phi_p} \sin\frac{\theta_p}{2} \cr
             e^{i\phi_p} \sin\frac{\theta_p}{2} & \cos \frac{\theta_p}{2}
          \end{pmatrix}.
\end{eqnarray}
\end{subequations}
The two bases are related by the equation, $\Psi_{p} = U_p \Phi_{p}$ for $p=L,R$. Obviously
the two bases at other sites are related to each other by the same unitary transformation.
Using the nonequilibrium Green's function method, the spin-polarized current can be expressed as
\begin{eqnarray}
\label{current_A}
I &=& \frac{2e}{h} \int d\epsilon ~ \mbox{Im} \mbox{Tr} V_{LR} G_{RL}^{<} (\epsilon), 
\end{eqnarray}
where the hopping matrix $V_{LR}$ and the Green's function are defined as
\begin{eqnarray}
V_{LR} &=& \begin{pmatrix} t_{L\uparrow,R\uparrow} & t_{L\uparrow,R\downarrow} \cr
                t_{L\downarrow,R\uparrow} & t_{L\downarrow,R\downarrow}
           \end{pmatrix},   \\
i\hbar G_{RL} (t,t') &=& \langle T \Psi_R (t) \Psi_L^{\dag} (t') \rangle.
\end{eqnarray}
Note that the hopping matrix has no dependence on the wave vector in the tight-binding 
Hamiltonian approach. 
Here $G_{RL} (t,t')$ is the time-ordered Green's function and $G_{RL}^{<}$ is the lesser
part of $G_{RL}$. After Fourier transform over time variable, the lesser Green's function 
$G_{RL}^{<}$ can be found in a closed form (see the main text and also [\onlinecite{neqgreen2}].)
\begin{eqnarray}
G_{RL}^{<} (\epsilon) 
 &=& \left[ 1 - G_{RL0}^{r} (\epsilon) V_{LR} \right]^{-1} G_{RL0}^{<} (\epsilon) \nonumber\\
 && \times \left[ 1 - V_{LR} G_{RL0}^{a} (\epsilon) \right]^{-1}, \\
G_{RL0}^{r,a} (\epsilon) 
 &=& G_{R}^{r,a} (\epsilon) V_{RL} G_{L}^{r,a} (\epsilon), \\
G_{RL0}^{<} (\epsilon) 
 &=& G_R^{<}(\epsilon) V_{RL} G_L^{a}(\epsilon) 
   + G_R^{r}(\epsilon) V_{RL} G_L^{<}(\epsilon).
\end{eqnarray}
Here $G_L$ and $G_R$ are the Green's functions at the left and right terminal sites, respectively,
when the coupling between two FM electrodes is absent.
\begin{eqnarray}
i\hbar G_p (t,t') &=& \langle T \Psi_p(t) \Psi_p^{\dag} (t') \rangle
  ~=~ U_p \langle T \Phi_p(t) \Phi_p^{\dag} (t') \rangle U_p^{\dag}, ~~~ p = L,R.
\end{eqnarray}
The superscripts $r,a$ and $<$ mean the retarded, advanced, and lesser part of the Green's function.
To compute the spin-polarized current, we have only to find the Green's functions at 
the terminal site ($L$ or $R$) when the coupling between two FM electrodes is absent. 
It is convenient to define the terminal Green's function in the spin diagonal basis
\begin{eqnarray}
i\hbar g_p (t,t') &=& \langle T \Phi_p(t) \Phi_p^{\dag} (t') \rangle
  ~=~ \begin{pmatrix} \langle T c_{p+}(t) c_{p+}^{\dag} (t') \rangle  & 0 \cr
           0  &  \langle T c_{p-}(t) c_{p-}^{\dag} (t') \rangle
      \end{pmatrix}.
\end{eqnarray}
This Green's function $g_p = \left( \begin{smallmatrix} g_{p+} & 0 \cr 
0 & g_{p-} \end{smallmatrix} \right)$ is diagonal and is related to $G_p$ 
by the unitary transformation, $G_p = U_p g_p U_p^{\dag}$.

 At this point we had better take a concrete example for the terminal Green's functions,
$g_L$ and $g_R$ (in the spin diagonal basis). 
As an example we consider the semi-infinite chain model for the FM electrodes.
It is relatively simple to find the self-energy of the terminal Green's function,
when we take into account the repeating structure of the chain. The self-energy can be expressed
in terms of the Green's function itself ($p=L,R$).
\begin{eqnarray}
\Sigma_p (t,t') 
  &=& \begin{pmatrix} |W_{p+}|^2 g_{p+}(t,t') & 0 \cr 0 & |W_{p-}|^2 g_{p-}(t,t') \end{pmatrix}. 
\end{eqnarray} 
From the Dyson equation, we can find the desired Green's functions in a straightforward manner.
The retarded and advanced Green's functions are 
\begin{eqnarray}
\label{chaingreen_A}
g_{p\sigma}^{r,a} (\epsilon)
 &=& \frac{2} 
   {\epsilon - \epsilon_{p\sigma} \pm i \sqrt{4|W_{p\sigma}|^2 - (\epsilon - \epsilon_{p\sigma})^2}}.
\end{eqnarray}
The retarded part ($r$) corresponds to the $+$ sign, while the advanced part ($a$) to the $-$ sign.
Quite in general, the retarded and advanced Green's functions at the terminal site
can be written in a form
$g_{p\sigma}^{r} (\epsilon) = R_{p\sigma} (\epsilon) - i \pi N_{p\sigma} (\epsilon)$ 
and $g_{p\sigma}^{a} = [g_{p\sigma}^{r}]^*$, respectively, 
such that we can write the matrix Green's function as
\begin{eqnarray}
g_{p}^{r,a} (\epsilon) 
  &=& \begin{pmatrix} R_{p+} (\epsilon) & 0 \cr 0 & R_{p-} (\epsilon) 
      \end{pmatrix} 
  \mp i \pi \begin{pmatrix} N_{p+} (\epsilon) & 0 \cr 0 & N_{p-} (\epsilon) 
            \end{pmatrix}
  ~=~ R_p (\epsilon) \mp i \pi N_p (\epsilon).
\end{eqnarray}
Here $N_{p\sigma}$ is the local density of states (DOS) or the local spectral function 
at the terminal site in the spin state $\sigma = \pm$. 
The lesser Green's function at the terminal site is simply related to the DOS matrix 
$N_p = \left( \begin{smallmatrix} N_{p+} & 0 \cr 0 & N_{p-} \end{smallmatrix} \right)$ 
by the expression, $g_p^{<} = 2\pi f_p(\epsilon) N_p$.

 Once the terminal Green's functions are known in the diagonal basis, the algebra becomes
straightforward to reduce the expression of the spin-polarized current, Eq.~(\ref{current_A}). 
Then the desired Green's functions at the terminal sites are ($p=L,R$)
\begin{eqnarray}
G_p^{r,a} (\epsilon) 
 &=& U_p \begin{pmatrix} g_{p+}^{r,a} (\epsilon) & 0 \cr 0 & g_{p-}^{r,a} (\epsilon) 
         \end{pmatrix} U_p^{\dag},  \\
G_p^{<} (\epsilon) 
 &=& 2\pi f_p(\epsilon) U_p N_p U_p^{\dag}. 
\end{eqnarray} 
To simplify the notations we rewrite
\begin{eqnarray}
G_p^{r,a} (\epsilon) 
  &=& \hat{R}_p \mp i \pi \hat{N}_p,  \\
G_p^{<} (\epsilon) 
 &=& 2\pi f_p(\epsilon) \hat{N}_p,
\end{eqnarray}
where $\hat{N}_p = U_p N_p U_p^{\dag}$ and $\hat{R}_p = U_p R_p U_p^{\dag}$.
The expression of $G_{RL0}^{<}$ can be rearranged as
\begin{eqnarray}
G_{RL0}^{<} &=& 2i\pi^2 [ f_R - f_L ] \hat{N}_R T_{RL} \hat{N}_L \nonumber\\
  && + 2\pi \left[  f_R \hat{N}_R T_{RL} \hat{R}_L + f_L \hat{R}_R T_{RL} \hat{N}_L \right]. 
\end{eqnarray}

 From now on we consider the case of no spin-flip in tunneling of electrons between 
two FM electrodes or $T_{RL} = t_{RL} {\bf 1}$. 
We also confine our interest to the case that two FM electrodes are the same material. 
Then we can derive the simpler relation for the spin-polarized current.
\begin{eqnarray}
G_{RL0}^{<} &=& 2i t_{RL} \pi^2 [ f_R - f_L ] \hat{N}_R \hat{N}_L   \nonumber\\
  && + 2\pi t_{RL} \left[  f_R \hat{N}_R \hat{R}_L + f_L \hat{R}_R \hat{N}_L \right],   \\
\mbox{Tr} \{ T_{RL} G_{RL}^{<} \}
  &=& t_{LR} \mbox{Tr} \left\{ [ 1 - t_{LR} G_{RL0}^r ]^{-1} G_{RL0}^{<} \right.  \nonumber\\
  && \times  \left.   [ 1 - t_{LR} G_{RL0}^a ) ]^{-1} \right\}.
\end{eqnarray}
From these two relations, we find the simple expression for $\mbox{Im} \mbox{Tr} T_{LR} G_{RL}^{<}$
in the parallel and antiparallel alignments of magnetizations such that the expression 
of the spin-polarized current is 
\begin{eqnarray}
I &=& \frac{2e}{h} \int d\epsilon ~ [f_R(\epsilon) - f_L(\epsilon)] ~ T(\epsilon),  \\
T(\epsilon) 
  &=& 2\pi^2 |t_{LR}|^2 ~ \mbox{Tr} \left\{ [ 1 - t_{LR} G_{RL0}^r ]^{-1} \hat{N}_R \hat{N}_L \right.
       \nonumber\\
  &&  \left.   [ 1 - t_{LR} G_{RL0}^a ]^{-1}  \right\}.
\end{eqnarray}
Since we are considering two FM electrodes of the same material, 
we have for the parallel alignment of magnetizations
\begin{eqnarray}
\hat{N}_L &=& \hat{N}_R ~=~ \begin{pmatrix} N_{+} & 0 \cr 0 & N_{-} \end{pmatrix},  \\
G_{RL0}^{r,a} 
  &=& t_{RL} \begin{pmatrix} (R_{+} \mp i \pi N_{+})^2 & 0 \cr 
                 0 & (R_{-} \mp i \pi N_{-})^2  \end{pmatrix}. 
\end{eqnarray}
The transmission probability for this case is 
\begin{eqnarray}
T_P (\epsilon) 
  &=& \sum_{\sigma = \pm} \frac{ 2\pi^2 |t_{LR}|^2 N_{\sigma}^2 }
     { \left| 1 - [ t_{LR} g_{\sigma} ]^2  \right|^2 }, 
\end{eqnarray}
where $g_{\sigma} =  R_{\sigma} + i \pi N_{\sigma}$.
For the antiparallel alignment of magnetizations, we have
\begin{eqnarray}
\hat{N}_L &=& \begin{pmatrix} N_{+} & 0 \cr 0 & N_{-} \end{pmatrix}, ~~~
  \hat{N}_R ~=~ \begin{pmatrix} N_{-} & 0 \cr 0 & N_{+} \end{pmatrix},  \\
G_{RL0}^{r,a} 
  &=& t_{RL} ( R_{+} \mp i \pi N_{+} ) ( R_{-} \mp i \pi N_{-} ) ~ {\bf 1}, 
\end{eqnarray}
such that the transmission probability is given by the expression
\begin{eqnarray}
T_{AP} (\epsilon) 
  &=& \frac{ 4\pi^2 |t_{LR} |^2 N_{+} N_{-} }  
           { \left| 1 - |t_{LR}|^2 g_{+} g_{-} \right|^2 }. 
\end{eqnarray}
We are now in a position to discuss any possible inverse tunneling magnetoresistance effect
for the narrow channel magnetic tunnel junctions.

 In a weak tunneling limit or when $|t_{LR} g_{\sigma}^{r}| \ll 1$, the transmission probabilities
for the parallel and antiparallel alignments of magnetizations can be approximated as
\begin{eqnarray}
T_P &=& 2\pi^2 |t_{LR}|^2 \left[ N_{+}^2 + N_{-}^2 \right],  \\
T_{AP} &=& 4\pi^2 |t_{LR} |^2 N_{+} N_{-} .
\end{eqnarray}
Obviously $T_P > T_{AP}$ such that more current flows in P than in AP, and 
the positive TMR ratio is obtained with the familiar Julliere expression.

 In a strong tunneling limit, let us see if there is a possibility of $T_P < T_{AP}$.
To be concrete let us consider the semi-infinite chain model. 
Note that the Eq.~(\ref{chaingreen_A}) can be written as
\begin{eqnarray}
g_{p\sigma}^{r} (\epsilon)
 &=& \frac{1} {|W_{p\sigma}|} \left[ \frac{ \epsilon - \epsilon_{p\sigma} }{ 2 |W_{p\sigma}| }
    - i \sqrt{ 1 -  \frac{(\epsilon - \epsilon_{p\sigma})^2}{ 4 |W_{p\sigma}|^2 }  }  
    \right]
\end{eqnarray}
and $g_{p\sigma}^{a} = [g_{p\sigma}^{r}]^*$. Since we are interested in the linear response 
transport, we have only to consider the terminal Green's functions at the Fermi level 
or $\epsilon = 0$. In a wide band limit or when $|\epsilon_{p\sigma}| \ll |W_{p\sigma}|$, 
the terminal Green's functions are reduced to $g_{p\sigma}^{r} \approx -i/|W_{p\sigma}|$ and 
$g_{p\sigma}^{a} \approx i/|W_{p\sigma}|$ such that our discussion corresponds exactly to our study
in the main text.  
When $R_{\pm} = 0$, the transmission probabilities for the tight-binding model are 
exactly the same as those in the main text. It is expected that the inverse TMR is 
possible when $|R| \ll N$.

\end{widetext}


\begin{thebibliography}{100}

\bibitem{review} E. Y. Tsymbal, O. N. Mryasov, and P. R. LeClair, 
  J. Phys.: Condens. Matter {\bf 15}, 109 (2003).
\bibitem{review2} I. Zutic, J. Fabian, and S. D. Sarma, Rev. Mod. Phys. {\bf 76}, 323 (2004).
\bibitem{moodera} J. S. Moodera, L. R. Kinder, T. M. Wong, and R. Meservey, 
  Phys. Rev. Lett. {\bf 74}, 3273 (1995).
\bibitem{miyazaki} T. Miyazaki and N. J. Tezuka, 
  J. Magn. Magn. Mater. {\bf 139}, L231 (1995).

\bibitem{julliere} M. Julliere, Phys. Lett. {\bf 54A}, 225 (1975).

\bibitem{teresa1} J. M. De Teresa, A. Barth\'{e}l\'{e}my, A. Fert, J. P. Contour, R. Lyonnet, 
  F. Montaigne, P. Seneor, and A. Vaur\`{e}s, 
  Phys. Rev. Lett. {\bf 82}, 4288 (1999).
\bibitem{teresa2} J. M. De Teresa, A. Barth\'{e}l\'{e}my, A. Fert, J. P. Contour,  
  F. Montaigne, and P. Seneor, 
  Science {\bf 286}, 507 (1999).
\bibitem{sharma} M. Sharma, S. X. Wang, and J. H. Nickel, 
  Phys. Rev. Lett. {\bf 82}, 616 (1999).
\bibitem{itmr_tsymbal} E. Y. Tsymbal, A. Sokolov, I. F. Sabirianov, and B. Doudin, 
  Phys. Rev. Lett. {\bf 90}, 186602 (2003).
\bibitem{fdotf_exp} A. N. Pasupathy, R. C. Bialczak, J. Martinek, J. E. Grose, 
  L. A. K. Donev, P. L. McEuen, and D. C. Ralph, Science {\bf 306}, 86 (2004). 


\bibitem{sdos_tsymbal} E. Y. Tsymbal and D. G. Pettifor, 
  J. Phys.: Condens. Matter {\bf 9}, L411 (1997). 
\bibitem{sdos_wang} K. Wang, S. Zhang, P. M. Levy, L. Szunyogh and P. Weinberger, 
  J. Magn. Mag. Mater. {\bf 189}, L131 (1998).
\bibitem{sdos_butler} W. H. Butler, X.-G. Zhang, T. C. Schulthess, and J. M. MacLaren,
  Phys. Rev. B {\bf 63}, 054416 (2001).



\bibitem{fdotf_the} 
 P. Zhang, Q.-K. Xue, Y. Wang, and X. C. Xie, Phys. Rev. Lett. {\bf 89}, 286803 (2002); 
 J. Martinek, Y. Utsumi, H. Imamura, J. Barna\'{s}, S. Maekawa, J. K\"{o}nig,
  and G. Sch\"{o}n, {\it ibid} {\bf 91}, 127203 (2003); 
 M.-S. Choi, D. S\'{a}nchez, and R. L\'{o}pez, {\it ibid} {\bf 92}, 056601 (2004). 


\bibitem{slonczewski} J. C. Slonczewski, Phys. Rev. B {\bf 39}, 6995 (1989).
\bibitem{transfer_ham1} J. Bardeen, Phys. Rev. Lett. {\bf 6}, 57 (1961).
\bibitem{transfer_ham2} T. E. Feuchtwang, Phys. Rev. B {\bf 10}, 4121 (1974); 
   {\it ibid} {\bf 10}, 4135 (1974).
\bibitem{neqgreen} L. P. Kadanoff and G. Baym, {\it Quantum Statistical Mechanics} 
 (Benjamin, New York, 1962).
\bibitem{neqgreen1} L.V. Keldysh, Zh. \'{E}ksp. Teor. Fiz. {\bf 47}, 1515 (1964).
 [Sov. Phys. JETP {\bf 20}, 1018 (1965)].
\bibitem{neqgreen2} T.-S. Kim and S. Hershfield, Phys. Rev. B {\bf 63}, 245326 (2001).

\bibitem{neqgreen3} T.-S. Kim and S. Hershfield, Phys. Rev. Lett. {\bf 88}, 136601 (2002); 
   Phys. Rev. B {\bf 67}, 165313 (2003).


\bibitem{langreth}  D.C. Langreth, 
  1976, in {\it Linear and Nonlinear Electron Transport in Solids}, Vol.{\bf 17} of 
  {\it NATO Advanced Study Institute, Series B: Physicsi},
  edited by J.T. Devreese and V.E. van Doren (Plenum, New York, 1976), p. 3.

\bibitem{wideband} The meaning of a wide band limit is clarified in the Appendix. 
  See, for example, S. Hershfield, J. H. Davies, and J. W. Wilkins, 
  Phys. Rev. Lett. {\bf 67}, 3720 (1991).

\bibitem{spinband} See, for example, M. Zwolak and M. Di Ventra, Appl. Phys. Lett. {\bf 81},
 925 (2002). 

\end{thebibliography}
\end{document}